# Multi-Level Attention Aggregation for Language-Agnostic Speaker Replication

Yejin Jeon
GSAI POSTECH
jeonyj0612@postech.ac.kr

Gary Geunbae Lee
GSAI POSTECH
CSE POSTECH
gblee@postech.ac.kr

## Abstract

This paper explores the task of language-agnostic speaker replication, a novel endeavor that seeks to replicate a speaker's voice irrespective of the language they are speaking. Towards this end, we introduce a multi-level attention aggregation approach that systematically probes and amplifies various speaker-specific attributes in a hierarchical manner. Through rigorous evaluations across a wide range of scenarios including seen and unseen speakers conversing in seen and unseen lingua, we establish that our proposed model is able to achieve substantial speaker similarity, and is able to generalize to out-of-domain (OOD) cases.

## 1 Introduction

Recent years have witnessed significant advancements in speech synthesis research, with notable contributions from well-established models like Tacotron (Wang et al., 2017; Shen et al., 2018), FastSpeech (Ren et al., 2019, 2021) and VITS (Kim et al., 2021). These models have enabled the generation of natural-sounding speech, which has prompted a notable shift in TTS research towards the synthesis of speech in the voices of both seen and unseen speakers in the domain of multi-speaker TTS. However, despite the considerable body of work in speaker imitation, it has primarily operated within the constraints of either a single target language or a predetermined set of languages. Consequently, the concept of generating speech in the voice of any speaker, regardless of the language spoken by that individual, remains largely unexplored - a novel concept we introduce as language-agnostic speaker replication.

There are two primary research domains dedicated to synthesizing text into audio with a target speaker's voice: voice cloning and multi-speaker TTS. Voice cloning entails the alteration of a speaker's voice without changing the text, and typically relies on two audio inputs from the source and target speakers (Wang et al., 2023; Tang et al., 2022). Although high speaker fidelity is achieved, this task operates within strict textual constraints as just the speaker's identity is modified. In contrast, multi-speaker TTS research bypasses the linguistic constraints of voice cloning as it aims to generate speech for any text using a specific speaker's voice. This task requires two modalities during inference: the text to be spoken, and an reference audio of the target speaker's voice (Min et al., 2021; Karlapati et al., 2022). While the target text may differ entirely from the spoken contents within the target speaker's audio, it still requires that the language being spoken in the reference audio is identical with that of the target text as in voice cloning.

| Task | Text | Speaker | Speaker Language |
|---|---|---|---|
| Voice Cloning | P | S + U | S |
| Multi-speaker TTS | P + U | S + U | S |
| Cross-lingual TTS | P + U | S | S |
| Language-Agnostic TTS | P + U | S + U | S + U |

Table 1: Criteria for task classification: 1) Text: Parallel or unparallel alignment of spoken content in the reference audio and target text input, 2) Speaker: Ability to replicate seen or unseen speakers, 3) Reference Audio: Whether the spoken language in the reference audio can be seen or unseen.

Unlike the aforementioned tasks, cross-lingual TTS studies extend the capabilities of multi-lingual TTS systems by allowing them to generate speech in a specific speaker's voice for text in different languages. Yet, these endeavors predominantly revolve around addressing the limitations of training data, where each speaker typically speaks just one language (Zhang et al., 2019). Thus, the primary objective is to facilitate voice transfer across different languages, making in a sense, a speaker a polyglot. More importantly, cross-lingual TTS only adapts the voices of seen speakers to a predefined set of languages present within the training data (Nekvinda and Dušek, 2020; Piotrowski et al., 2023). On the other hand, the task of language-agnostic speaker replication goes beyond voice cloning, multi-speaker TTS, and cross-lingual TTS (Table 1); it focuses on accurate imitation of seen and unseen speakers' voices even when they speak in an unseen language, while keeping the language of the target text fixed. This task holds significant potential for various applications including movie dubbing, and other scenarios where voice preservation is vital.

In light of these considerations, we pose the following questions: Can we achieve high speaker fidelity even when the reference audio's spoken *language* differs from the target text, and is not represented in the training data? Moreover, can this be done in a zero shot manner?

Our contributions in pursuit of this novel challenge encompass three aspects: 1) we advance zero-shot multi-speaker TTS with language-agnostic speaker imitation, 2) investigate a multi-level attention aggregation approach for enhancing speaker fidelity, and 3) demonstrate the effectiveness of our methodology through comparative analyses with baseline models while ensuring a thorough and diverse evaluation by conducting validations for eleven different languages across eight phylogenetic language branches.

## 2 Methodology

### 2.1 Preliminaries

Our model is composed of three main components: 1) an acoustic model based on Ren et al. (2021), 2) a speaker module, and 3) a HiFi-Gan (Kong et al., 2020) vocoder. In this section, we explicitly focus on the speaker module, which is instrumental in modeling speaker information from a reference audio in a language-agnostic fashion. We employ SALN (Min et al., 2021) to merge the final speaker embedding from the speaker module with the acoustic model.

### 2.2 Multi-Level Attention Aggregation

In order to extract language-agnostic speaker representations for conditioning the acoustic model, we begin by utilizing an ECAPA-TDNN (Desplanques et al., 2020; Ravanelli et al., 2021) speaker verification (SV) model pretrained on the VoxCeleb datasets (Nagrani et al., 2020; Chung et al., 2018). Given a variable-length audio sequence $X = [x_1, x_2, ..., x_n]$, this input is passed through the SV model, which is made up of three 1-dim Squeeze-Excitation Res2Blocks each with a scale dimension of 8, and a channel dependent statistics pooling layer. This results in the generation of an intermediate speaker representation $z \in R^d$.

However, it is important to recognize that exclusive reliance on a pretrained speaker extractor as an immediate conditioning factor for acoustic modeling as in prior research (Jia et al., 2018; Xue et al., 2022) can prove to be inadequate, especially when dealing with OOD scenarios with unseen speakers and languages. To address this, we probe and amplify speaker-dependent properties within $z$ through multi-level attention aggregation.

Two main factors used to distinguish different speakers are fundamental frequency (F0) and timbre (Skuk et al., 2020). We first extract F0 contours using the Yin algorithm (de Cheveigné and Kawahara, 2002; Guyot, 2019) from $X$, which are then used to identify the congruence with intermediate representation $z$. In other words, if attributes in $z$ are confirmed by corresponding F0 values, those speaker features should be accentuated. Thus, we prompt $z$ with F0, which is formally organized as

$$H_{CA1} = softmax(\frac{q(H_{SV})k(H_{F0})}{\sqrt{d_k}})v(H_{F0}), \quad (1)$$

where $H_{SV}$ and $H_{F0}$ represent the states for $z$ and encoded F0 information, respectively.

While F0 serves as a local feature, timbre represents the global representation of a speaker's spectral envelope. Thus, we further investigate the interrelationships between these two distinct aspects of speaker information. Global information is first extracted from $X$ as mel-spectrograms using Short-Time Fourier transform (STFT, Griffin and Lim (1984)) of filter lengths 1024, window size 1024, hop size 256, and 80 frequency bins on input audio $X$. The resulting 2D speech representation then undergoes a series of transformations, including fully connected blocks, and a gated convolution block (Dauphin et al., 2017), which ultimately results in $H_{ME}$. To establish meaningful connection between the F0-accentuated representation $H_{CA1}$ and the global representation $H_{ME}$, we conduct probing with cross-attention. This process can be denoted as

$$H_{CA2} = softmax(\frac{q(H_{ME})k(H_{CA1})}{\sqrt{d_k}})v(H_{CA1}) \quad (2)$$

Furthermore, for the purpose of improving generalization capabilities, we introduce an additional step involving representation splitting. The representation derived from the second attention aggregation stage $H_{CA2}$ is partitioned into a set of randomly initialized $N$ embeddings. We then apply multi-head attention to ascertain the contribution of each embedding (Wang et al., 2018). The weights derived from this process are used to compute a weighted sum of the $N$ representations, resulting in the ultimate speaker representation for the input audio $X$. Losses are identical to Ren et al. (2021).

| | | Phylogeny | Language | MOS (↑) | | | ABX (↑) | | |
|---|---|---|---|---|---|---|---|---|---|
| | | | | Baseline 1 | Baseline 2 | Proposed | Baseline 1 | Baseline 2 | Proposed |
| Seen Languages | Seen Speakers | Western Romance | Portuguese | 3.12 ± 0.14 | 2.70 ± 0.14 | 3.52 ± 0.09 | 0.36 | 0.24 | 0.40 |
| | | West Slavic | Polish | 3.04 ± 0.09 | 2.80 ± 0.11 | 3.62 ± 0.08 | 0.32 | 0.24 | 0.44 |
| | | Koreanic | Korean | 3.18 ± 0.10 | 3.06 ± 0.14 | 3.48 ± 0.08 | 0.36 | 0.24 | 0.40 |
| | | West Germanic | English | 3.02 ± 0.14 | 3.46 ± 0.11 | 3.44 ± 0.09 | 0.28 | 0.40 | 0.32 |
| | Unseen Speakers | Indo-Aryan | Hindi | 2.72 ± 0.14 | 2.92 ± 0.14 | 3.42 ± 0.13 | 0.32 | 0.28 | 0.40 |
| | | West Germanic | English | 2.94 ± 0.16 | 3.36 ± 0.11 | 3.24 ± 0.14 | 0.36 | 0.32 | 0.32 |
| | | Western Romance | Spanish | 2.68 ± 0.13 | 2.90 ± 0.13 | 3.58 ± 0.14 | 0.16 | 0.20 | 0.64 |
| | | Western Romance | French | 3.14 ± 0.12 | 2.92 ± 0.12 | 3.56 ± 0.10 | 0.44 | 0.04 | 0.52 |
| Unseen Languages | Unseen Speakers | North Germanic | Icelandic | 3.18 ± 0.12 | 2.98 ± 0.12 | 3.64 ± 0.11 | 0.36 | 0.24 | 0.40 |
| | | Southern Bantu | Xhosa | 3.06 ± 0.13 | 3.38 ± 0.12 | 3.18 ± 0.09 | 0.32 | 0.40 | 0.28 |
| | | Malayo-Polynesian | Javanese | 3.38 ± 0.12 | 2.86 ± 0.14 | 3.22 ± 0.12 | 0.52 | 0.12 | 0.36 |
| | | West Germanic | Afrikaans | 2.96 ± 0.11 | 3.06 ± 0.14 | 3.63 ± 0.11 | 0.24 | 0.32 | 0.44 |

Table 2: MOS with 95% confidence intervals, and ABX results. Xue et al. (2022), and x-vector (Snyder et al., 2018) conditioned FastSpeech2 are referred to as Baselines 1 and 2, respectively.

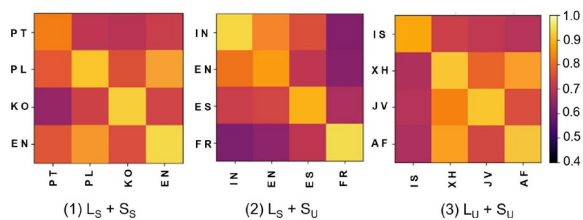

Figure 1: Intra-group cross-similarity matrices, arranged from left to right as "Seen Language Seen Speaker ($L_S + S_S$)," "Seen Language Unseen Speaker ($L_S + S_U$)," and "Unseen Language Unseen Speakers ($L_U + S_U$)." Higher speaker similarities between utterances are represented by brighter, yellow hues.

## 3 Experiments

We leverage the train-clean-100 subset of the LibriTTS dataset (Zen et al., 2019) to train the entire TTS system in an end-to-end manner, which encompasses 53 hours of English recordings distributed among 247 speakers of nearly balanced gender ratio. Audios are sampled at 22050 Hz and 16 bits. All computational models were trained on a single RTX A6000 GPU for approximately 20 hours until step 300,000. We use $\beta_1 = 0.9$, $\beta_2 = 0.98$, and $\epsilon = 10^{-9}$ hyperparameters for Adam optimization. Our model has a total of 27,056,339 parameters.

## 4 Results and Discussion

### 4.1 Evaluation Protocol

We employ MOS and ABX evaluations to assess the quality of our synthesized speech (Appendix A). These evaluations were conducted on the Amazon Mechanical Turk platform with 25 participants. During the MOS evaluations, participants independently compared each synthesized speech to the ground truth reference audio, providing ratings for speaker fidelity on a Likert scale ranging from 1 to 5, with 0.5 increments. On the other hand, the ABX evaluations required participants to select only one synthetic audio out of multiple choices that exhibited the closest speaker similarity to the reference audio. The average duration for these assessments was approximately 40 minutes.

### 4.2 Assessment Analyses

We conducted a comparative analysis between our proposed model and the baselines Xue et al. (2022) and x-vector (Snyder et al., 2018) adapted to Ren et al. (2021) (Table 2). Higher MOS and ABX scores reflects the proposed model's ability to better generalize across a diverse group of speakers, even when they are speaking in different languages, while demonstrating higher speaker fidelity in synthetic speech. To further substantiate our findings, we performed cross similarity calculations (Figure 1). This involved the comparison of twelve utterances from different speakers with twelve other utterances from the same set of speakers. The highest similarity between utterances is consistently observed when they are synthesized with the same speaker, underscoring the model's efficacy in preserving the speaker's distinctive characteristics.

To investigate the contribution of individual attributes at different attention aggregation levels, we conducted ablation studies (Table 3). In the initial stage preceding multi-level attention aggregation, we exclusively utilized $H_{SV}$ representations. While some improvements were observed compared to baseline models, they were not particularly significant. Instead, MOS scores saw notable increases with first-level attention aggregation involving interactions between $H_{SV}$ and either $H_{ME}$ or $H_{F0}$. While there was a higher preference for using $H_{F0}$ as the initial prompt for $H_{SV}$, the two models generally exhibited complementary outcomes. These results thus reinforce the importance of incorporating both local and global speaker attributes.

| | | Seen Language Seen Speaker | | | | Seen Language Unseen Speaker | | | | Unseen Language Unseen Speaker | | | |
|---|---|---|---|---|---|---|---|---|---|---|---|---|---|
| Level | Model | Portuguese | Polish | Korean | English | Hindi | English | Spanish | French | Icelandic | Xhosa | Javanese | Afrikaans |
| 0 | SE | 3.18 ± 0.15 (0.16) | 2.86 ± 0.14 (0.12) | 2.76 ± 0.17 (0.08) | 3.14 ± 0.14 (0.12) | 2.94 ± 0.18 (0.08) | 2.74 ± 0.18 (0.08) | 2.94 ± 0.17 (0.12) | 2.66 ± 0.18 (0.04) | 3.14 ± 0.16 (0.08) | 2.78 ± 0.17 (0.52) | 2.88 ± 0.16 (0.64) | 2.94 ± 0.15 (0.40) |
| 1 | SE + ME | 2.96 ± 0.14 (0.12) | 3.06 ± 0.14 (0.24) | 3.20 ± 0.11 (0.12) | 3.10 ± 0.13 (0.20) | 3.16 ± 0.16 (0.20) | 3.32 ± 0.12 (0.24) | 3.20 ± 0.13 (0.20) | 3.44 ± 0.11 (0.20) | 3.34 ± 0.13 (0.24) | 3.34 ± 0.12 (0.32) | 3.22 ± 0.11 (0.24) | 3.12 ± 0.10 (0.16) |
| 1 | SE + F0 | 3.04 ± 0.09 (0.20) | 2.66 ± 0.07 (0.08) | 3.32 ± 0.08 (0.20) | 3.22 ± 0.10 (0.24) | 3.20 ± 0.10 (0.24) | 3.14 ± 0.11 (0.28) | 3.06 ± 0.14 (0.16) | 3.48 ± 0.08 (0.20) | 3.28 ± 0.11 (0.20) | 3.20 ± 0.12 (0.20) | 3.08 ± 0.09 (0.16) | 3.16 ± 0.09 (0.20) |
| 2 | (SE + ME) + F0 | 2.88 ± 0.14 (0.08) | 2.72 ± 0.17 (0.16) | 2.92 ± 0.14 (0.12) | 3.08 ± 0.13 (0.12) | 2.90 ± 0.16 (0.12) | 2.98 ± 0.14 (0.08) | 2.98 ± 0.12 (0.12) | 3.10 ± 0.14 (0.16) | 3.24 ± 0.13 (0.12) | 3.08 ± 0.12 (0.12) | 3.00 ± 0.13 (0.16) | 2.82 ± 0.13 (0.04) |
| 2 | (SE + F0) + ME | 3.48 ± 0.11 (0.36) | 3.40 ± 0.09 (0.32) | 3.62 ± 0.10 (0.40) | 3.46 ± 0.10 (0.36) | 3.36 ± 0.12 (0.32) | 3.38 ± 0.11 (0.44) | 3.50 ± 0.11 (0.36) | 3.64 ± 0.13 (0.36) | 3.30 ± 0.11 (0.28) | 3.28 ± 0.11 (0.24) | 3.48 ± 0.09 (0.40) | 3.66 ± 0.10 (0.52) |

Table 3: Ablation results for multi-level attention aggregation. ABX scores are indicated within parentheses below their respective MOS results. SE, ME, and F0 are the equivalents of $H_{SV}$, $H_{ME}$, and $H_{F0}$, respectively.

Subsequently, we experimented with integrating the remaining component that was not used in the preceding attention aggregation step. Adding global information at a later stage (i.e., (SE + F0) + ME), consolidated in notably higher scores for both MOS and ABX. This supports previous findings that the aggregation of $H_{SV}$ and $H_{F0}$ has higher speaker fidelity compared to $H_{SV}$ and $H_{ME}$ aggregation. In summary, it is evident that the combination of both local and global speaker information is crucial, and initialization via fundamental frequency prompting leads to better speaker imitation.

We further validate these findings by visualizing synthetic audio generated for all models using identical text and reference audios (Figure 2). Notably, the SE model, which does not employ any attention aggregation, exhibited the lowest and incorrect pitch contours, along with distortion in the high formant frequencies, which are associated with gender identification (Poon and Ng, 2015). Introduction of either local or global information resulted in an overall increase in pitch levels and reduced high formant frequency distortion. Yet, the most accurate pitch contours and formants were obtained with two-step attention aggregation using $H_{F0}$ interpolation before $H_{ME}$. Moreover, when comparing the proposed (SE + F0) + ME model with and without representation splitting, the latter exhibited overall similar speaker similarity, albeit with a reduction in local pitch fluctuation. This suggests that while representation splitting may not substantially affect speaker fidelity, it contributes to enhanced intonation and overall naturalness.

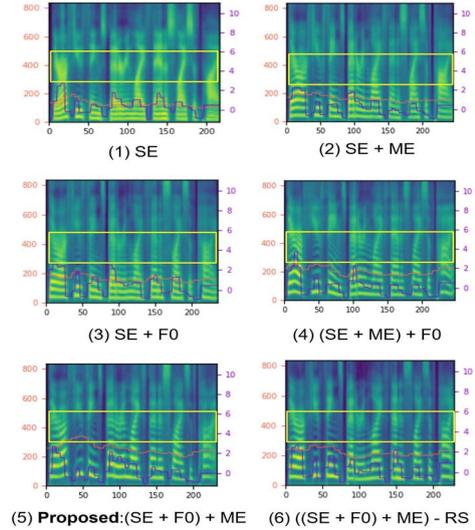

Figure 2: Pitch (orange), energy (purple), formant (yellow box) visualizations for each model in Table 3. Identical text and a female speaker's reference audio were used for synthesis. RS refers to replication splitting.

## 5 Conclusion

In this study, we have formulated the novel task of language-agnostic speaker replication, presented a multi-level attention aggregation approach tailored for this task, and explored the intricate interplay of diverse speaker attributes. We have also effectively addressed the fundamental questions we initially posed: 1) Even when the language of the reference audio differs from the target text and is unrepresented in the data, high speaker fidelity is achieved. 2) Zero-shot speaker replication for both seen and unseen speakers is possible, demonstrating the robustness and versatility of our approach. We anticipate that our preliminary efforts will serve as a foundation for further developments in this domain.

# 6 Limitations

In this work, we made several contributions: 1) We defined the task of language-agnostic speaker replication, 2) proposed a novel multi-level attention aggregation method for this task, and 3) conducted comprehensive evaluations spanning multiple languages. Our primary focus was on the language-agnostic speaker extraction aspect within the broader TTS framework. Consequently, we did not extend our current work to include a multilingual context, which would entail synthesizing text inputs in multiple languages, akin to multilingual TTS. Thus, we intend to undertake this as part of our future research objectives.

Given the nature of this work, it is imperative to recognize the potential ramifications of unethical utilization such as identity misappropriation and coercive inducement of actions contrary to an individual's consent. We acknowledge these potential risks, and are fully committed to responsible research and usage of voice replication technology.

## Acknowledgements

This work was supported by the Technology Innovation Program(20015007, Development of Digital Therapeutics of Cognitive Behavioral Therapy for treating Panic Disorder) funded By the Ministry of Trade, Industry & Energy(MOTIE, Korea), and by Smart HealthCare Program(www.kipot.or.kr) funded by the Korean National Police Agency(KNPA, Korea) [Project Name: Development of an Intelligent Big Data Integrated Platform for Police Officers' Personalized Healthcare / Project Number: 220222M01].

## A  Reference Audio Datasets

Our evaluations encompass three distinct groups of speakers: 1) Seen Languages Seen Speakers, 2) Seen Languages Unseen Speakers, and 3) Unseen Languages Unseen Speakers. The first group encompasses speakers and languages that were part of the dataset employed for pretraining the ECAPA-TDNN SV model. In the second category, we utilize speakers whose spoken language is found within the dataset, but the speakers themselves are not. The third group includes speakers and languages that are entirely absent from the dataset, and originate from diverse open-source datasets (van Niekerk et al., 2017; Mollberg et al., 2020; Sodimana et al., 2018).

## B  Crowdsourcing for Model Assessments

We employ the Amazon Mechanical Turk platform to enlist participants for our MOS and ABX evaluations. Our instructions to participants emphasize their exclusive focus on the assessment of voice similarity among two or more audio samples. It is important to note that we do not request any personal information from the participants in this process, ensuring their privacy and security. To determine a fair and adequate payment structure, we initially conducted a pilot test with one individual. Following the completion of the pilot test within a one-hour time frame, we concluded that compensating participants according to the hourly wage of the author's country was a reasonable and equitable approach.